\documentclass[epjCONF]{svjour}
\usepackage{graphics}
\usepackage[varg]{txfonts} 
\usepackage[latin1]{inputenc}
\session-title{Conference Title, to be filled}
\begin{document}
\title{Extracting surface rotation periods of solar-like \emph{Kepler} targets}
\author{T. Ceillier\inst{1}\fnmsep\thanks{\email{tugdual.ceillier@cea.fr}} \and R.A. Garc\'ia\inst{1} \and D. Salabert\inst{1} \and S. Mathur\inst{2}}
\institute{Laboratoire AIM, CEA/DSM/IRFU/SAp - CNRS - Univ. Paris Diderot, Centre de Saclay, 91191 Gif-sur-Yvette Cedex, France
 \and Space Science Institute, 4750 Walnut Street, Suite 205, Boulder, Colorado 80301 USA}
\abstract{
We use various method to extract surface rotation periods of \emph{Kepler} targets exhibiting solar-like oscillations and compare their results.
} 
\maketitle
\section{Introduction}
\label{Intro}
Rotation plays a very important part in the evolution of a star, mainly because of transport processes due to meridional circulation. Furthermore, it remains difficult to account for the the internal rotation profiles of sub and red giants that were inferred from asteroseismic studies. In this work, we study the surface rotation rates and activity levels of \emph{Kepler} solar-like oscillating stars - including Main-Sequence stars, subgiants and red giants - which are good asteroseimic targets.

\section{Methodology}
\label{Methodo}

In order to get a robust determination of rotation periods, we use two types of data correction -- PDC-MAP \cite{ThompsonRel21} and KADACS \cite{2011MNRAS.414L...6G} -- as well as two different ways of getting an estimation of the rotation period -- wavelets decomposition like in \cite{2010A&A...511A..46M} and autocorrelation function following \cite{2013MNRAS.432.1203M}. We then compare the four results obtained. If they all concur, we select the period obtained as the rotation period with a high confidence. As a last verification, we check visually all the stars for which a rotation period has been derived. The activity index $\langle S_{ph,k=5} \rangle$ is then calculated as in \cite{2014JSWSC...4A..15M}.

The results of this methodology applied on the sample of solar-like oscillating stars on the Main Sequence and the Subgiant phases have been reported in \cite{2014arXiv1403.7155G}.

\section{Example for a red giant star}
\label{RG}

This methodology has also been used to study a sample of \emph{Kepler} Red Giants and will be described in details in Ceillier et al. 2015 (in preparation). Due to their low activity levels, these stars are not supposed to show clear rotational modulations in their lightcurve. This is why only a small fraction of the global sample (around 2\%) give conclusive results. These peculiar Red Giants could result from mergers, as discussed by Tayar et al. 2015 (in preparation).

\begin{figure}
\resizebox{0.92\columnwidth}{!}{
  \includegraphics{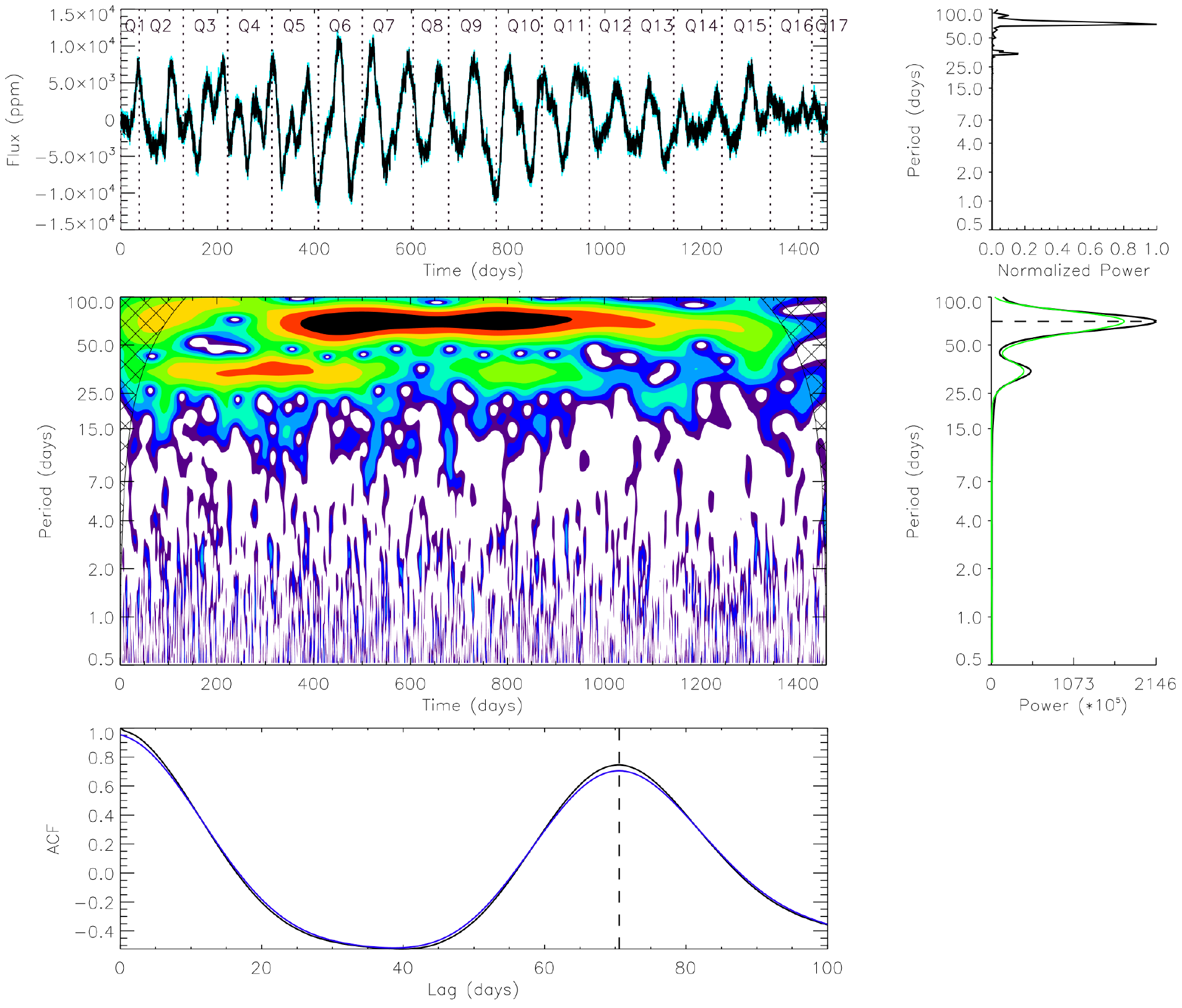} }
\caption{Example of the analysis for KIC~2570214 (KADAC data). Top panel: Long-cadence \emph{Kepler} light curve (cyan) and rebinned light curve (black), where vertical dotted lines indicate the transitions between the observing quarters. Top right panel: associated power density spectrum as a function of period between 0.5 and 100 days. Middle left panel: Wavelet Power Spectrum (WPS) computed using a Morlet wavelet between 0.5 and 100 days on a logarithmic scale. The black-crossed area is the cone of influence corresponding to the unreliable results. Middle right panel: Global Wavelet Power Spectrum (GWPS) as a function of the period of the wavelet and the associated fit composed from several gaussian functions (thin green line). The horizontal dashed line designates the position of the retrieved $P_{\rm{rot}}$. Bottom panel: AutoCorrelation Function (ACF) of the full light curve plotted between 0 and 100 days (black) and smoothed ACF (blue). The vertical dashed line indicates the returned  $P_{\rm{rot}}$ for the ACF analysis.}
\label{ceillier:fig1}       
\end{figure}

An example of a Reg Giant star's light curve analysis can be seen in Fig.~\ref{ceillier:fig1}. The 70~days modulation is clearly visible.

\bibliographystyle{epj}  
\bibliography{Ceillier_T.bib}

\begin{thebibliography}{6}

\bibitem{ThompsonRel21}
S.E. Thompson, J.L. Christiansen, J.M. Jenkins, D.A. Caldwell, T.~Barclay, S.T.
  Bryson, C.J. Burke, J.~Campbell, B.D. Catanzarite, B.D. Clarke et~al.,
  \emph{{Kepler Data Release 21 Notes (KSCI-19061-001)}}, Kepler mission (2013)

\bibitem{2011MNRAS.414L...6G}
R.A. Garc\'{\i}a, S.~Hekker, D.~Stello, J.~Guti\'{e}rrez-Soto, R.~Handberg,
  D.~Huber, C.~Karoff, K.~Uytterhoeven, T.~Appourchaux, W.J. Chaplin et~al.,
  Monthly Notices of the RAS \textbf{414}, L6 (2011), \texttt{1103.0382}

\bibitem{2010A&A...511A..46M}
S.~Mathur, R.A. Garc\'{\i}a, C.~R\'{e}gulo, O.L. Creevey, J.~Ballot,
  D.~Salabert, T.~Arentoft, P.O. Quirion, W.J. Chaplin, H.~Kjeldsen, Astronomy
  and Astrophysics \textbf{511}, A46 (2010), \texttt{0912.3367}

\bibitem{2013MNRAS.432.1203M}
A.~McQuillan, S.~Aigrain, T.~Mazeh, Monthly Notices of the RAS \textbf{432},
  1203 (2013), \texttt{1303.6787}

\bibitem{2014JSWSC...4A..15M}
S.~Mathur, D.~Salabert, R.A. Garc\'{\i}a, T.~Ceillier, Journal of Space Weather
  and Space Climate \textbf{4}, A15 (2014), \texttt{1404.3076}

\bibitem{2014arXiv1403.7155G}
R.A. Garc\'{\i}a, T.~Ceillier, D.~Salabert, S.~Mathur, J.L. van Saders,
  M.~Pinsonneault, J.~Ballot, P.G. Beck, S.~Bloemen, T.L. Campante et~al.,
  ArXiv e-prints  (2014), \texttt{1403.7155}

\end{thebibliography}

\end{document}